\documentstyle[preprint,aps]{revtex}
\begin{document}

\title{
\begin{flushright}
\vspace{-1cm}
{\normalsize MC/TH 96/19}
\vspace{1cm}
\end{flushright}
Determination of pion-baryon coupling constants\\
 from QCD sum rules}
\author{Michael C. Birse and Boris Krippa\thanks{Permanent address: Institute 
for Nuclear Research of the Russian Academy of Sciences, Moscow Region 117312,
Russia.}}
\address{Theoretical Physics Group, Department of Physics and Astronomy\\
University of Manchester, Manchester, M13 9PL, UK\\}
\maketitle
\begin{abstract}
We evaluate the $\pi NN$, $\pi\Sigma\Sigma$ and $\pi\Sigma\Lambda$ coupling
constants using QCD sum rules based on pion-to-vacuum matrix elements of
correlators of two interpolating baryon fields. The parts of the correlators
with Dirac structure $k\llap/\gamma_5$ are used, keeping all terms up to
dimension 5 in the OPE and including continuum contributions on the
phenomenological side. The ratios of these sum rules to baryon mass sum rules
yield stable results with values for the couplings of $g_{\pi NN}=12\pm 5$,
$g_{\pi\Sigma\Sigma}=7\pm 4$ and $g_{\pi\Sigma\Lambda}=6\pm 3$. The
sources of uncertainty are discussed.
\end{abstract}
\bigskip
\section{Introduction}

Meson-baryon coupling constants form an important ingredient in many
calculations of strong-interaction processes and one would like to determine
these quantities from QCD. In the absence of treatments from first principles,
the method of QCD sum rules\cite{svz79} has proved to be a very powerful tool
for studying various properties of low-lying hadron states. Here we apply this
method to the calculation of the coupling constants of pions to the lowest
states of the baryon octet: $N$, $\Lambda$ and $\Sigma$.

The pion-nucleon coupling constant $g_{\pi NN}$ has previously been studied
within the framework of QCD sum rules by several groups
\cite{rry83,rry85,sh95,bk96}. Reinders, Rubinstein and Yazaki\cite{rry85}
explored two different approaches, one based on the correlator of three
interpolating fields sandwiched between vacuum states, and one based on the
pion-to-vacuum matrix element of the correlator of two interpolating nucleon
fields, $\eta$:
\begin{equation}
\langle0|T\{\eta(x)\overline{\eta}(0)\}|\pi^a(k)\rangle,
\label{twopt}
\end{equation}
The particular sum rule they studied was based on the soft-pion limit of the
the part of two-point correlator (1) with Dirac structure $\gamma_5$. However
those authors took into account only the leading term of the operator product
expansion (OPE) and they neglected continuum contributions. Shiomi and
Hatsuda\cite{sh95} extended the analysis of this sum rule to include
condensates up to dimension 7 in the OPE as well as a perturbative estimate of 
continuum contributions. 

The sum rules that we use here are also constructed from two-point
correlators (\ref{twopt}) of the appropriate baryon interpolating fields. The
advantage of this method is that it allows one to calculate hadron properties
at low values of the momentum transfer to the baryon. In contrast, the
straightforward use of OPE for the three-point correlator is valid only for
large spacelike meson momenta and therefore a determination of the coupling
constant requires an extrapolation to zero momentum where OPE is clearly not
valid because of large power corrections. Estimates of the coupling constant
 from the coefficient of $1/k^2$ determined at large $k^2$, as in
Refs.\cite{rry83,rry85,mei95}, cannot distinguish the pole term of lowest
meson from the contributions of higher-mass states in the same meson channel
with the same $1/k^2$ behavior at large $k^2$.

We note that modified versions of the OPE of three-point correlators for the
processes with small momentum transfer have been developed in
Refs.\cite{by83,is84}. The essence of these methods is the inclusion of
``bilocal power corrections," which effectively sum up the series of power
terms in $1/k^2$ by matching them to the contributions of mesonic states in
the relevant channel. The contributions of low-lying mesons to the form
factors play an increasingly important role as the momentum transfer
decreases. Meson-baryon coupling constants can be obtained from the OPE of
three-point correlator with bilocal power corrections by going to the meson
pole. At the pole this treatment of the three-point correlator yields the same
results as the method based on two-point correlator which is used in this
paper (cf.\cite{bbkr95}).

The particular sum rules that we study here are constructed from the part of
the correlator (\ref{twopt}) with Dirac structure $k\llap/\gamma_5$. We chose
this structure because it provides a determination of the pion-baryon
couplings that is not simply related to sum rules for the baryon masses. In
contrast the soft-pion limit of the OPE for the $\gamma_5$ piece of the
two-point correlator for $g_{\pi NN}$ has exactly the same form as that for
the nucleon sum rule\cite{iof81,is84} involving condensates of odd dimension,
up a factor of $1/f_\pi$ \cite{rry85,sh95}. Shiomi and Hatsuda\cite{sh95}
showed that the ratio of the $\gamma_5$ sum rule to one for the nucleon mass
takes the form of the Goldberger-Treiman relation with $g_A$=1, provided that
continuum thresholds are taken to be the same in both cases. Those authors
took different thresholds in the two sum rules in order to to get around this
problem with the implied value of $g_A$.

However, we stress that taking soft-pion limit of the $\gamma_5$ piece of the
two-point correlator (\ref{twopt}) does not lead to an independent
determination of the coupling constant. In the case of $g_{\pi NN}$,
the usual soft-pion theorem\cite{dgh92}, can be used to express the correlator 
(\ref{twopt}) in the form
\begin{equation}
-{i\over f_\pi}\langle 0|[Q_5^a, T(\eta(x),\eta^\dagger(0))]|0\rangle
={i\over 2f_\pi}\{\gamma_5\tau^a,\langle 0|T(\eta(x),
\eta^\dagger(0))|0\rangle\}
\end{equation}
where $Q_5^a$ is the axial charge and we have made use of the transformation
properties of the interpolating field under axial rotations\cite{el90,lccg95},
$[Q_5^a,\eta]=-{1\over 2}\gamma_5\tau^a\eta$. The anticommutator with
$\gamma_5$ picks out the part of the two-point correlator proportional to the
unit Dirac matrix. The phenomenological side of the resulting sum rule is thus
$i\gamma_5/f_\pi$ times the corresponding expression for the odd-condensate
nucleon sum rule. This matches exactly with the structure found for the OPE
side in Refs.\cite{rry85,sh95}.

The soft-pion limit for the $\gamma_5$ piece of the correlator (1) thus yields
a sum rule for $M_N/f_\pi=g_{\pi NN}/g_A$. The value for the coupling
determined from such a rum rule follows from the odd-condensate sum rule for
the nucleon mass and the Goldberger-Treiman relation (or an approximation to
it taking $g_A=1$). The sum rule can be thought of as just a chiral rotation
of the odd-condensate nucleon sum rule and {\it not} an independent
determination of $g_{\pi NN}$. Physically this result is quite natural since 
in the soft-pion limit $\pi B$ and $B$ states become degenerate and can be
related to each other by chiral transformations. In this paper, by considering
terms beyond the soft-pion limit, we obtain values for pion-baryon couplings
that are not simply consequences of chiral symmetry.

In addition we note that a potentially important piece of the phenomenological
side is missing from previous sum-rule determinations of $g_{\pi NN}$. This
term corresponds to transitions of where a ground-state baryon created by the
interpolating field absorbs the pion and is excited into the continuum. Since
they are not suppressed by the Borel transformation such terms should be
included in a consistent sum-rule analysis, as pointed out long
ago\cite{is84,bk84} and stressed recently by Ioffe\cite{iof95a,iof95b}. In the
soft-pion limit of the $\gamma_5$ sum rule, such terms generate contact
interactions where the pion couples directly to the baryon field, $\langle
B(p)|\overline{\eta}_{n}(0)|\pi(k)\rangle$, and which are essential if the
correct soft-pion limit is to be obtained. The omission of these terms in
Refs.\cite{rry85,sh95} can explain why the correct Goldberger-Treiman relation
was not found there. Indeed, as the authors of\cite{sh95} point out, a quick
estimate of these unsuppressed $N^*$ contributions suggests that they could be
as large as 25\%: enough to remove the discrepancy with the Goldberger-Treiman
relation.

As discussed above, the sum rules studied here provide values for the
pion-baryon couplings that are not simply related to the baryon masses by
chiral symmetry. We include all condensates up to dimension 5 as well as
mixed continuum terms. These are essential for assessing the reliability of
the sum rules and estimating the uncertainties in the results. The application
of these sum rules to $g_{\pi NN}$ has been described briefly in\cite{bk96}.
Similar sum rules have been applied to other pion couplings, especially in the
context of $D$ and $B$ mesons, as discussed in\cite{bbkr95} and references
therein.

The paper is organised as follows: in Sec.\ II we derive sum rules for the
pion-baryon couplings from the relevant two-point correlators; the numerical
analysis of the sum rules is presented in Sec.\ III; finally our results are
summarized in Sec.\ IV.

\section{Two-point correlators and sum rules}

Our sum rules are obtained from the two-point correlator (\ref{twopt}) just
discussed, but instead of the piece with with Dirac structure $\gamma_5$
considered in Refs.\cite{rry85,sh95} we work with the structure
$k\llap/\gamma_5$, where $k$ is the pion momentum. We work here to leading
order in a chiral expansion, neglecting higher-order terms in the pion
momentum or current quark mass. To illustrate the derivation of sum rules for
pion-baryon couplings, we consider first the sum rule for $g_{\pi NN}$. The
differences that arise for the pion-hyperon couplings will then be discussed
and the forms of the resulting sum rules presented.

We consider the two-point correlation function
\begin{equation}
\Pi(p)=i\int d^4x\exp(ip\cdot x)\langle 0|T\{\eta_{p}(x)
\overline{\eta}_{n}(0)\}|\pi^+(k)\rangle,    
\label{correl}
\end{equation}
where we use the Ioffe interpolating field\cite{iof81} for the proton,
\begin{equation}
\eta_{p}(x)=\epsilon_{abc}[u^{a}(x)^TC\gamma_{\mu}u^{b}(x)]\gamma_{5}
\gamma^{\mu}d^{c}(x),
\label{ioffep}
\end{equation}
and the corresponding neutron field $\eta_n$ which is obtained by
interchanging $u$ and $d$ quark fields. Here $a,b,c$ are the colour indices
and $C$ is the charge conjugation matrix. Other choices of interpolating field 
can be used, as discussed in detail by Leinweber\cite{lei95}. For the
odd-condensate nucleon sum rule, which we make use of in our determination of
$g_{\pi NN}$, it turns out that the Ioffe field is close to optimal\cite{lei95} 
and so we do not consider more general fields.

In the deeply Euclidean region, where $p^2$ is large and negative, the OPE of 
the product of two interpolating fields takes the following general form
\begin{equation}
i\int d^4x\exp(ip\cdot x)T\{\eta_{p}(x)\overline{\eta}_{n}(0)\}
=\sum_n C_n(p)O_{n},
\end{equation}
where $C_{n}(p)$ are the Wilson coefficients and $O_{n}$ are local operators
constructed out of quark and gluon fields (all renormalised at some scale
$\mu$). Using this OPE in correlators of the form (\ref{correl}), we find that
only operators of odd dimension contribute. The leading term in this expansion
involves operators with dimension 3 and is given by
\begin{equation}
\Pi_{3}(p,k)=-\frac{1}{2\pi^{2}}p^{2}\ln(-p^{2})\langle 0|\overline{d}
\gamma^{\alpha}\gamma_{5}u|\pi^+(k)\rangle\gamma_{\alpha}\gamma_{5}+\cdots,
\end{equation}
where terms that do not contribute to the Dirac structure of interest, 
$k\llap/\gamma_5$, have been suppressed. The matrix element here is just 
the usual one for pion decay:
\begin{equation}
\langle 0|\overline{d}\gamma^{\alpha}\gamma_{5}u|\pi^+(k)\rangle=i\sqrt{2}
f_\pi k^\alpha,
\end{equation}
where $f_\pi=93$ MeV is the pion decay constant. Hence we can write the
leading term as
\begin{equation}
\Pi_{3}(p,k)=-i\sqrt{2}\frac{1}{2\pi^{2}}p^{2}\ln(-p^{2})f_\pi k\llap/
\gamma_{5}+\cdots,
\label{dim3}
\end{equation}

At dimension 5 the only relevant contribution arises from the second-order term
in the covariant expansion of the nonlocal operator $\overline{d}(0)
\gamma^{\alpha}\gamma_{5}u(x)$. This is a specific feature of the Ioffe
nucleon interpolating field\cite{iof81} which we used to calculate $\Pi^{N}$.
This term has the form
\begin{equation}
\Pi_{5}(p)=\frac{5}{9\pi^{2}}\ln(-p^{2})\langle 0|\overline{d}\gamma^{\alpha}
\gamma_{5}D^{2}u|\pi^+(k)\rangle\gamma_{\alpha}\gamma_{5}+\cdots.
\label{dim5ex}
\end{equation}
Up to corrections of higher order in the current mass, the matrix element here
can easily be re-expressed in terms of a mixed quark-gluon condensate
\begin{equation}
\langle 0|\overline{d}\gamma^{\alpha}\gamma_{5}D^{2}u|\pi^+(k)\rangle=
\frac{g_{s}}{2}\langle 0|\overline{d}\gamma^{\alpha}\gamma_{5}\sigma_{\mu\nu}
G^{\mu\nu}u|\pi^+(k)\rangle+{\cal O}(m_c^2).
\end{equation}
With some further manipulation this can be rewritten in the form
\begin{equation}
\langle 0|\overline{d}\gamma^{\alpha}\gamma_{5}D^{2}u|\pi^+(k)\rangle=
-g_{s}\big(\langle 0|\overline{d}\widetilde{G}^{\alpha\mu}\gamma_{\mu}u
|\pi^+(k)\rangle-ig_s\langle 0|\overline{d}G^{\mu\alpha}\gamma_{\mu}\gamma_{5}u
|\pi^+(k)\rangle\bigr),
\label{cond5}
\end{equation}
where $\widetilde{G}_{\mu\nu}=\frac{1}{2}\epsilon_{\mu\nu\rho\sigma}
G^{\rho\sigma}$. (We use the convention $\epsilon^{0123}=+1$.) The second term
in this expression is of higher order in the chiral expansion (see
Ref.\cite{nsvvz} for details) and so we neglect it here.

The first term in (\ref{cond5}) is of leading order in the chiral expansion.
It involves a matrix element that has been extracted by Novikov {\it et
al.}\cite{nsvvz} from two QCD sum rules for the pion. They expressed it in the
form
\begin{equation}
g_{s}\langle 0|\overline{d}\widetilde{G}^{\alpha\mu}\gamma_{\mu}u
|\pi^+(k)\rangle=\sqrt{2}i\delta^{2}f_{\pi}k^{\alpha},
\label{delta2}
\end{equation}
and obtained $\delta^2=(0.20\pm 0.02)$ GeV$^2$. In both their sum rules
the four-quark condensate, $\alpha_s\langle 0|(\overline qq)^2 |0 \rangle$
makes a crucial contribution. Novikov {\it et al.}\cite{nsvvz} used the
factorisation approximation for this quantity in their analysis. However
direct determinations of it from other sum rules lead to
values\cite{kar92,bnp92,nar95} that are at least 2--3 times bigger than those
obtained from factorisation. These give correspondingly larger values for
$\delta^2$, a point we shall come back to in the analysis of the sum rules in
Sec.~III. Our final expression for the dimension-5 term in the sum rule is
\begin{equation}
\Pi_{5}(p)=-i\sqrt{2}\frac{5}{9\pi^{2}}\ln(-p^{2})\delta^2f_\pi k\llap/
\gamma_{5}+\cdots.
\label{dim5}
\end{equation}

To estimate of the importance of higher dimension condensates, we have also
calculated the contribution of what we hope is the most important dimension-7
operator in the OPE. This is a mixed quark-gluon condensate, which we evaluate
in the factorised approximation. Keeping only this contribution explicitly,
the dimension-7 piece of the correlator is
\begin{equation}
\Pi_{7}(p)=-\frac{1}{12p^{2}}\langle 0|\overline{d}\gamma^{\alpha}\gamma_{5}u
|\pi^+(k)\rangle\langle 0|\frac{\alpha_{s}}{\pi}G^{2}|0\rangle\gamma_{\alpha}
\gamma_{5}+\cdots,
\label{dim7}
\end{equation}
where $\langle 0|\frac{\alpha_{s}}{\pi}G^{2}|0\rangle$ is the gluon condensate
in vacuum. We find that the contribution of this condensate is small, as
discussed in the following section.

On the phenomenological side, the $\pi N$ coupling constant is contained in the
term of the correlator (\ref{correl}) with a double pole at the nucleon mass.
However there are also continuum contributions which cannot be ignored. 
These include continuum-to-continuum pieces which can be
modelled in the usual manner, in terms of the spectral density associated with
the imaginary part of the OPE expression for the correlator. This continuum is
assumed to start at some threshold $S_{\pi N}$. After Borel transformation, it
can be taken over to the OPE side of the sum rule where it modifies the
coefficients of the terms involving $\ln(-p^2)$. In addition one must include
nucleon-to-continuum terms since Borel transformation does not suppress these
with respect to the double-pole term\cite{is84,bk84,iof95a,iof95b}. To first
order in $k$, the correlator has the form
\begin{equation}
\Pi(p)=i{\sqrt 2}k\llap/\gamma_5\left[{\lambda_N^2 M_Ng_{\pi NN}\over 
(p^2-M_N^2)^2}+\int_{W^2}^\infty ds\,b(s){1\over s-M_N^2}
\left({1\over p^2-M_N^2}+{a(s)\over s-p^2}\right)\right]+\cdots,
\label{phenom}
\end{equation}
where the continuum-continuum terms (and terms with other Dirac structures)
have not been written out. Here $\lambda_N$ is the strength with which the
interpolating field couples to the nucleon:
\begin{equation}
\langle 0|\eta_N(0)|N(p)\rangle=\lambda_N u({\hbox{p}}).
\end{equation}

The sum rule is obtained by equating the OPE and phenomenological expressions 
for the correlator (\ref{correl}) and Borel transforming\cite{svz79}. Keeping
only condensates up to dimension 5, this has the form
\begin{equation}
\frac {1}{2\pi^{2}}M^4E_{2}(x_{\pi N}) + \frac{5}{9\pi^{2}}M^2E_{1}(x_{\pi N})
\delta^{2}
=\left({\lambda_N^2 M_N g_{\pi NN}\over f_{\pi}M^2}+A\right)
\exp(-M^{2}_{N}/M^{2}),
\label{gpinsr}
\end{equation}
where $M$ is the Borel mass and $E_{n}(x) = 1 - (1 + x +...+\frac{x^{n}}{n!})
e^{-x}$ with $x_{\pi N} = \frac{S_{\pi N}}{M^{2}}$. The second term on the
r.h.s.~of this sum rule is the Borel transform of the nucleon pole term of the
nucleon-to-continuum piece in (\ref{phenom}). It involves an undetermined
constant $A$ but, since it contains the same exponential as the nucleon
double-pole term, it cannot be ignored. The second nucleon-to-continuum term
in (\ref{phenom}) leads to a term that is suppressed by an exponential
involving the masses of states in the continuum. It is thus typically a factor
of 3--4 smaller than the term included in (\ref{gpinsr}). Provided that the
first of these mixed terms is a reasonably small correction to the sum rule,
it should be safe to neglect the second, as discussed by
Ioffe\cite{iof95a,iof95b}.

The construction of sum rules for the pion-hyperon couplings follows similar 
lines. For the $\Sigma^{+,0}$ and $\Lambda$ we use the following 
fields, obtained by SU(3) rotations of (\ref{ioffep})\cite{iof81}:
\begin{equation}
\eta_{\Sigma^{+}}(x)=\epsilon_{abc}[u^{a}(x)^TC\gamma_{\mu}u^{b}(x)]\gamma_{5}
\gamma^{\mu}s^{c}(x),
\end{equation}
\begin{equation}
\eta_{\Sigma^{0}}(x)=\sqrt{2}(\eta_{Y2}(x)+\eta_{Y1}(x)),
\end{equation}
\begin{equation}
\eta_{\Lambda}(x)=\sqrt{2\over 3}(\eta_{Y2}(x)-\eta_{Y1}(x)),
\end{equation}
where we have introduced
\begin{equation}
\eta_{Y1}(x)=\epsilon_{abc}[d^{a}(x)^TC\gamma_{\mu}s^{b}(x)]\gamma_{5}
\gamma^{\mu}u^{c}(x),
\end{equation}
\begin{equation}
\eta_{Y2}(x)=\epsilon_{abc}[u^{a}(x)^TC\gamma_{\mu}s^{b}(x)]
\gamma_{5}\gamma^{\mu}d^{c}(x),
\end{equation}

It is convenient to evaluate the correlators of $\eta_{Y1}$ and $\eta_{Y2}$
with the $\Sigma^+$ field separately. Considering $\eta_{Y1}$ first. we find 
that its correlator has the same basic form as the proton-neutron one just
discussed. The only difference is that it is smaller by a factor of two since
it contains only one strange-quark field. For the $k\llap/\gamma_5$ piece of
this correlator we therefore have
\begin{equation}
\Pi^{Y1}(p)=-i\sqrt{2}\frac{1}{4\pi^{2}}p^{2}\ln(-p^{2})f_\pi k\llap/
\gamma_{5}
-i\sqrt{2}\frac{5}{18\pi^{2}}\ln(-p^{2})\delta^2f_\pi k\llap/
\gamma_{5}+\cdots.
\label{y1corr}
\end{equation}

The OPE for the correlator of $\eta_{Y2}$ starts with a dimension-3 term of
the form
\begin{equation}
\Pi^{Y2}_3(p)=i\sqrt 2\frac{1}{24\pi^2}p^2\ln(-p^2)f_\pi k\llap/
\gamma_5+\cdots.
\label{y2dim3}
\end{equation}
Unlike the corresponding terms in (\ref{dim3},\ref{y1corr}), which have the
form $k\llap/\gamma_5/x^6$ in coordinate space, this term arises from one of
the form $x\llap/ x\cdot k/x^8$. This difference in the coordinate-space
structure means that the corresponding dimension-5 term coming from the
expansion of $\overline{d}(0)\gamma^{\alpha}\gamma_{5}u(x)$ has a different
relative coefficient compared to that in (\ref{dim5},\ref{y1corr}). It involves
the same matrix element (\ref{delta2}) discussed above and has the form
\begin{equation}
\Pi^{Y2}_5(p)=i\sqrt{2}\frac{5}{72\pi^{2}}\ln(-p^{2})\delta^2f_\pi k\llap/
\gamma_{5}+\cdots.
\end{equation}
One might have expected an additional contribution of this form from the
background gluon field in the quark propagator. However it turns out that such
a term vanishes for the $k\llap/\gamma_5$ piece of the correlator of
$\eta_{Y2}$ and $\overline\eta_{\Sigma^+}$ because of a cancellation of
contributions from the coordinate-space forms $k\llap/\gamma_5/x^4$ and
$x\llap/ x\cdot k/x^6$.

At dimension 7 there are mixed quark-gluon condensate terms, which are similar
to the term in the the nucleon correlator (\ref{dim7}). The first
SU(3)-breaking term also appears at this order. This involves a condensate of
the form $m_s\langle 0|\overline{q}q\overline{d}\gamma^{\alpha}\gamma_{5}u
|\pi^+(k)\rangle$, stemming from the mass term in the strange-quark
propagator. The term can be estimated in the factorisation approximation and
we find that it gives a very small (less than 5\%) contribution to the OPE
side of the sum rules. We therefore neglect it in our analyses.

The phenomenological expressions for the hyperon correlators are
\begin{equation}
\Pi^\Sigma(p)=i k\llap/\gamma_5{\lambda_\Sigma^2 M_\Sigma g_{\pi \Sigma\Sigma}
\over (p^2-M_\Sigma^2)^2}+\cdots,
\end{equation}
\begin{equation}
\Pi^\Lambda(p)=-i k\llap/\gamma_5{\lambda_\Sigma \lambda_\Lambda
M_Yg_{\pi \Sigma\Lambda}\over 
(p^2-M_Y^2)^2}+\cdots,
\end{equation}
where only the pole terms have been written out. In the $\Lambda\Sigma$
correlator $M_Y$ denotes the average hyperon mass since we neglect the
mass difference between the $\Sigma$ and $\Lambda$. (The numerical
coefficients in the definitions of the coupling constants can be found
in\cite{ccrev}.)

Taking the combinations of the $\eta_{Y1}$ and $\eta_{Y2}$ correlators that
correspond to the $\Sigma^0$ and $\Lambda$ and equating them to the
phenomenological expressions, we obtain the sum rules
\begin{eqnarray}
\frac{5}{12\pi^{2}}M^4E_{2}(x_{\pi\Sigma}) + \frac{5}{12\pi^{2}}M^2
E_{1}(x_{\pi\Sigma})\delta^{2}
&=&\left({\lambda_{\Sigma}^2 M_{\Sigma}g_{\pi\Sigma\Sigma}\over f_{\pi}M^2}
+A_\Sigma\right)\exp(-M^{2}_{\Sigma}/M^{2}),\label{gpissr}\\
\frac{7}{12\pi^{2}}M^4E_{2}(x_{\pi\Lambda}) + \frac{25}{36\pi^{2}}M^2
E_{1}(x_{\pi\Lambda})\delta^{2}
&=&\sqrt{3}\left({\lambda_{\Sigma}\lambda_{\Lambda} M_Y g_{\pi\Sigma\Lambda}
\over f_{\pi}M^2}+A_\Lambda\right)\exp(-M^{2}_Y/M^{2}).\label{gpilsr}
\end{eqnarray}

In the limit of exact SU(3) symmetry there two independent couplings
of pseudoscalar mesons to the baryon octet, usually denoted $F$ and $D$
corresponding to antisymmetric and symmetric combinations of the octet fields.
The $\pi N$ coupling is proportional to $F+D$ and the hyperon couplings can
be written as 
\begin{eqnarray}
g_{\pi\Sigma\Sigma}&=&2\alpha g_{\pi NN},\label{su3s}\\
g_{\pi\Sigma\Lambda}&=&{2\over\sqrt{3}}(1-\alpha)g_{\pi NN},
\label{su3l}
\end{eqnarray}
where
\begin{equation}
\alpha={F\over F+D},
\end{equation}
(see, for example:\cite{nijm,juel}). Comparing our sum rules
(\ref{gpissr},\ref{gpilsr}) with these forms we see that, if the strengths
$\lambda_B$ are SU(3) symmetric, the correlator of $\eta_{Y1}$ contributes to
the coupling $F+D$, while $\eta_{Y2}$ contributes to $F-D$. In this limit
the dimension-3 terms in these sum rules would lead to an $F/D$ ratio of $5/7$,
although the dimension-5 terms would tend to reduce this value. For comparison, 
SU(6) quark models give $F/D=2/3$ and SU(3)-symmetric analyses of pion-baryon
couplings\cite{nijm,juel} or baryon axial couplings\cite{cr93} tend to give
values around 0.58. One should remember that SU(3) is significantly broken by
the strange quark mass and so it may not be possible to represent the couplings
in terms of $F$ and $D$.

\section{Analysis}

We now turn to the numerical analysis of these sum rules. First, one should get
rid of the unknown constants $A_B$. Multiplying the sum rules by
$M^2\exp{M_N^2/M^2}$, we see that the right-hand sides become linear functions
of $M^2$. By acting on these forms of the sum rules with
$(1-M^2\partial/\partial M^2)$\cite{is84} (or equivalently by fitting a
straight line to the left-hand sides and extrapolating to $M^2=0$\cite{bk84})
we can in principle determine value for the couplings. However we are unable
to find a region of Borel mass in which the left-hand sides are approximately
linear functions of $M^2$, and hence there is no region of stability for the
extracted $g_{\pi BB}$.

This lack of stability is similar to the situation for the nucleon sum rules,
where two sum rules can be derived\cite{iof81} (involving either odd or even
dimension operators) but neither shows good stability. Nonetheless the ratio of
these leads to a more stable expression for the nucleon mass. We have therefore
taken the ratio of our sum rules to those for the corresponding baryons. We
obtain the most stable results from the ratios to the following baryon sum
rules\cite{iof81,is84,rry85} (see also: \cite{fgc92,jf94,jn95}),
\begin{equation}
-{1\over 4\pi^2}M^4E_1(x_N)\langle 0|\overline q q|0\rangle+{1\over 24}
\langle 0|\overline q q|0\rangle\langle 0|\frac{\alpha_{s}}{\pi}G^{2}|0\rangle
=\lambda_N^2 M_N \exp(-M^{2}_{N}/M^{2}),
\label{noddsr}
\end{equation}
\begin{equation}
{m_{s}\over 16\pi^4}M^6E_2(x_\Sigma)-{1\over 4\pi^2}M^4E_1(x_\Sigma)\langle 0|
\overline s s|0\rangle+{4\over 3}m_s\langle 0|(\overline qq)^2|0\rangle
=\lambda_{\Sigma}^2 M_{\Sigma} \exp(-M^{2}_{\Sigma}/M^{2}),
\label{soddsr}
\end{equation}
\begin{eqnarray}
-{m_{s}\over 48\pi^4}M^6E_2(x_\Lambda)&-&{M^4\over 12\pi^2}(4\langle 0|
\overline q q|0\rangle-\langle 0|\overline ss|0\rangle) E_1(x_\Lambda) 
\label{loddsr}\\
&+&{4\over 9}m_s[3\langle 0|(\overline qq)^2|0\rangle-\langle 0|(\overline qq)
(\overline ss)|0\rangle]
=M_{\Lambda}\lambda_{\Lambda}^2\exp(-M^{2}_{\Lambda}/M^{2}),\nonumber
\end{eqnarray}
and so we present here only the results for these cases. Taking such ratios
also has the advantage of eliminating the experimentally undetermined strengths
$\lambda_B$ from the sum rules. Note that we have allowed for a different
continuum threshold $S_B$ in each of the sum rules and have defined
$x_B=S_B/M^2$.

Again we describe first the sum rule for $g_{\pi NN}$ and then discuss the
additional features that arise for the hyperons. We take the ratio of the sum
rules (\ref{gpinsr}) and (\ref{noddsr})
\begin{equation}
f_\pi{\frac {1}{2\pi^{2}}M^6E_{2}(x) + \frac{5}{9\pi^{2}}M^4E_{1}(x)\delta^{2}
+\frac{1}{12}M^2E_{0}(x)\langle 0|\frac{\alpha_{s}}{\pi}G^{2}|0\rangle\over
-{1\over 4\pi^2}M^4E_1(x_N)\langle 0|\overline q q|0\rangle+{1\over 24}
\langle 0|\overline q q|0\rangle\langle 0|\frac{\alpha_{s}}{\pi}G^{2}|0\rangle}
=g_{\pi NN}+A_N'M^2,
\label{ratio}
\end{equation}
and use the method discussed above to eliminate the unknown mixed
nucleon-to-continuum term, $A_N'M^2$ (where $A_N'=A_Nf_\pi/\lambda_N^2 M_N$).
The results for $g_{\pi NN}$ are shown in Fig.~1 as a function of the Borel
mass $M^2$. These have been obtained using the following typical values of the
condensates and thresholds:
$\langle 0|\overline q q|0\rangle=-(0.245\ {\hbox{GeV}})^3$,
$\langle 0|\frac{\alpha_{s}}{\pi}G^{2}|0\rangle\simeq 0.012$ GeV$^{4}$,
$\delta^2=0.35$ GeV$^2$, $S_N=2.5$GeV$^2$, and $S_{\pi N}=2.15$ GeV$^2$. Stable
values of $g_{\pi NN}\simeq 11.7$ are found over a region $M^2\simeq 0.8-1.8$
GeV$^2$. Corrections due to the $A_N'M^2$ term are small, at most 5\%. The
second such term in (\ref{phenom}) is expected to be smaller by a factor of
3--4, and so we are justified in neglecting it.

The threshold $S_{\pi N}$ has been adjusted so that stable results are
obtained for Borel masses around 1 GeV$^2$, since one may hope that in this
region the Borel transformed sum rule is not too sensitive to the
approximations that have been made on both the OPE and phenomenological sides
of the sum rule. The existence of a window of stability provides a check on
the consistency of this assumption. We also demand that the thresholds $S_N$
and $S_{\pi N}$ should lie significantly above this window so that the
continuum is not too heavily weighted in the Borel transform. We find that the
window of stability moves rapidly upwards as $S_{\pi N}$ is increased for
fixed $S_N$. For the typical parameter values above, only the region 2.05
GeV$^2\leq S_{\pi N}\leq 2.22$ GeV$^2$ satisfies these requirements. The value
of $g_{\pi NN}$ varies by at most $\pm 0.2$ over this region.

We have examined the dependence of our results to the threshold in the nucleon
sum rule $S_N$. Varying this from 2.2 to 2.8 GeV$^2$, readjusting $S_{\pi N}$
to maintain stability, changes $g_{\pi NN}$ by $\pm 0.2$. To estimate the
sensitivity of our sum rules to the contributions of dimension-7 condensates
and to uncertainties in the gluon condensate, we have varied the dimension-7
term in (\ref{gpinsr}) between zero and twice its standard value. Our results
for $g_{\pi NN}$ change by $\pm 0.5$ over this range.

As a further check on our results, we have examined whether the individual sum
rules (\ref{gpinsr}) and (\ref{noddsr}) satisfy the criteria suggested by
Leinweber\cite{lei95}. We find that the highest dimension condensates
contribute less that 10\% of the OPE to both sum rules for $M^2>0.8$ GeV$^2$.
The procedure of differentiation with respect to $M^2$ does tend to increase
the size of the continuum contribution. Nonetheless it does remain within
Leinweber's limit, forming about 40\% of the phenomenological side of the
differentiated version of the sum rule (\ref{gpinsr}) for $M^2$ up to 1.4
GeV$^2$, the point at which the continuum reaches 50\% of the odd-condensate
sum rule (\ref{noddsr}). We therefore use the region $M^2\simeq 0.8-1.4$
GeV$^2$ since this provides a window within which our results are both stable
with respect to the Borel mass and not too sensitive to our approximations.

We have also examined the dependence of our results on the other input
parameters. One of the most important of these is the matrix element
$\delta^2$, defined by (\ref{delta2}). As already mentioned, this parameter
was extracted by Novikov {\it et al.}\cite{nsvvz} from an analysis of two sum
rules for the pion. Their results depend crucially on the four-quark
condensate, $\alpha_s\langle 0|(\overline qq)^2 |0 \rangle$, for which they
made the factorisation approximation and took a value of about $2\times
10^{-4}$ GeV$^6$. With this input, both of their sum rules yield consistent
results for $\delta^2$ in the region $0.20 \pm 0.02$ GeV$^2$. However,
sum-rule analyses of $\tau$ decay and $e^+e^-$ annihilation into hadrons lead
to significantly larger values of the four-quark condensate
(see\cite{kar92,bnp92,nar95} and references therein), in the range
$(4-6)\times 10^{-4}$ GeV$^6$. Using these in the sum rules of
Ref.\cite{nsvvz} leads to values for $\delta^2$ ranging from 0.28 to 0.45,
although the two sum rules do not then give consistent results. As a
conservative estimate of the uncertainty in $\delta^2$ we have considered the
range 0.20 to 0.45 GeV$^2$. The corresponding variation in $g_{\pi NN}$ is $\pm
2$ when the other parameters  are held at their values above and $S_{\pi N}$ is
changed to keep the window of stability around 1 GeV$^2$.

The second significant source of uncertainty is the quark condensate 
$\langle 0|\overline q q|0\rangle$ which appears in the odd-dimension sum
rule for the nucleon. ``Standard" values for this lie in the range $-(0.21\
{\hbox{GeV}})^3$ and $-(0.26\ {\hbox{GeV}})^3$. The values of the baryon masses
determined from sum rules\cite{iof81} are strongly correlated with this
condensate. There is also a weaker correlation with the chosen value of the
threshold $S_B$. Since we are dividing our sum rules by baryon sum rules, our
results are rather sensitive to the value of this condensate. One would like
to use values of $\langle 0|\overline q q|0\rangle$ and $S_N$ that give, for
example the nucleon mass correctly, but the ratio of the odd and even
dimension nucleon sum rules does not yield completely stable results for
$M_N$. The best we can do is to rule out values of $-\langle 0|\overline q
q|0\rangle$ below $(0.23\ {\hbox{GeV}})^3$ since they cannot reproduce the
nucleon mass within the region of Borel mass and threshold that we consider.
Varying the quark condensate between $-(0.23\ {\hbox{GeV}})^3$ and $-(0.26\
{\hbox{GeV}})^3$, we find that $g_{\pi NN}$ changes by $\pm 2$.

Including all of these sources of uncertainty, our final result for the
pion-nucleon coupling constant is thus $g_{\pi NN}=12\pm 5$, where the error is
dominated by $\delta^2$ and $\langle 0|\overline q q|0\rangle$. This value is
to be compared with those deduced from $NN$ and $\pi N$ scattering. For many
years the accepted value was $g_{\pi NN}=13.4$\cite{bcc73} but this coupling
has been the subject of some debate in recent years. More recent analyses lead
to values in the range 12.7--13.6\cite{newgpi}. Our result is obviously
consistent with any of these. 

The analysis of the pion-hyperon sum rules follows similar lines. In these
cases additional input parameters are needed to describe the effects of SU(3)
breaking in the hyperon mass sum rules (\ref{soddsr}, \ref{loddsr}). For the
strange quark mass, we consider values in the range $m_s=130$--230
MeV\cite{gl82}. We write the strange quark condensate in the form $\langle
0|\overline s s|0\rangle$=$\gamma\langle 0|\overline q q|0\rangle$ and
consider $\gamma$ in the range 0.7--0.9. To allow for deviations from the
factorisation approximation, we write the four-quark condensates in the form
$\langle 0|(\overline q q)^2|0\rangle=K (\langle 0|\overline q q|0\rangle)^2$
and vary $K$ between 1 and 2.

For the $g_{\pi\Sigma\Sigma}$ sum rule we find a similar window of Borel
stability for values of $S_{\pi\Sigma}$ in the region 1.8 to 2 GeV$^2$,
provided we take $S_\Sigma$ in the range 2.8 to 3.0 GeV$^2$. With the typical
values for the parameters above and $m_s=180$ MeV, $\gamma=0.7$ and $K=1$, we
get $g_{\pi\Sigma\Sigma}\simeq 6.8$. The relative uncertainties in this
arising from $\delta^2$ and the quark condensate are similar to those for
$g_{\pi NN}$. There are also significant further uncertainties from $m_s$,
$\gamma$ and $K$, which add another $\pm 1$. Our final result for this
coupling is $g_{\pi\Sigma\Sigma}=7\pm 4$. A similar analysis for the
$g_{\pi\Lambda\Sigma}$ sum rule leads to $g_{\pi\Lambda\Sigma}=6\pm 3$. We
should also point out that there is an additional uncertainty in our
determination of the latter coupling since we have ignored the
$\Sigma$-$\Lambda$ mass splitting in obtaining the sum rule (\ref{gpilsr}).

Within our large error bars, these results for the pion-hyperon couplings are
compatible with the empirical values quoted in Ref.\cite{ccrev},
$g_{\pi\Sigma\Sigma}=13\pm 2$ and $g_{\pi\Lambda\Sigma}=12\pm 2$, as well as
more recent determinations\cite{nijm,juel}, which yield values in the range
10--12 for both couplings. However one should note that Refs.\cite{nijm,juel}
assume SU(3) symmetry of the couplings whereas our results show significant
SU(3) breaking and cannot be expressed in terms of $F$ and $D$ couplings.

The rather large uncertainties in these results could be reduced if the quark
condensate could be determined more precisely. In addition, the sum rules of
Novikov {\it et al.}\cite{nsvvz} should be re-examined using larger values of
the four-quark condensate to try to pin down the value of $\delta^2$ more
exactly. We also note that there are correlations amongst the parameters used,
for example between $\delta^2$ and the four-quark condensate, and so we may
have overestimated the total uncertainties to some extent. It might therefore
be worth applying the techniques of Leinweber\cite{lei95} to these sum rules.
However we note that recent applications of that approach to sum rules for the
axial coupling also lead to results with $\sim 50\%$ uncertainties\cite{llj96}.

\section{Summary}

We have calculated the pion-nucleon and pion-hyperon coupling constants using
QCD sum rules based on the pion-to-vacuum matrix element of a two-point
correlator of interpolating baryon fields. We have included
baryon-to-continuum terms omitted from previous analyses. Our sum rules are
based on the part of the correlator with Dirac structure $k\llap/\gamma_5$ and
includes all terms up to dimension 5 in the OPE. Stable results are obtained
 from the ratio of these sum rules to ones for the baryon masses and the
unsuppressed baryon-to-continuum contributions are found to be small.
Contributions from higher-dimension operators and omitted continuum terms are
estimated to be small. Within admittedly rather large errors, our results
for the coupling constants are consistent with the empirical values. 

One should note that the uncertainties in our results are large. While
we have indicated ways in which one might hope to reduce some of these
uncertainties, our results and those of\cite{llj96} for $g_A$ indicate that
sum rules for baryon couplings are unlikely ever to reach similar accuracy to
those for baryon masses. Nonetheless this approach may be able to yield useful
information on other couplings whose values are at present not well determined.

\section*{Acknowledgements}
We are grateful to V. Kartvelishvili and J. McGovern for useful discussions.
M.C.B.~thanks the TQHN group at the University of Maryland for its hospitality
during the completion of this work. This work was supported by the EPSRC and
PPARC.

\newpage
\begin{center}
{\large FIGURE CAPTION}
\end{center}
\bigskip
Fig.~1. Dependence on the square of the Borel mass of the $\pi NN$ coupling
constant determined from the ratio of sum rules for $M_{N}$ and $g_{\pi NN}$.
The values of the parameters used are given in the text. The solid line shows
the value of $g_{\pi NN}$ corrected for the mixed continuum term $A_N'M^2$, the
dashed line corresponds the uncorrected value of $g_{\pi NN}$.
 
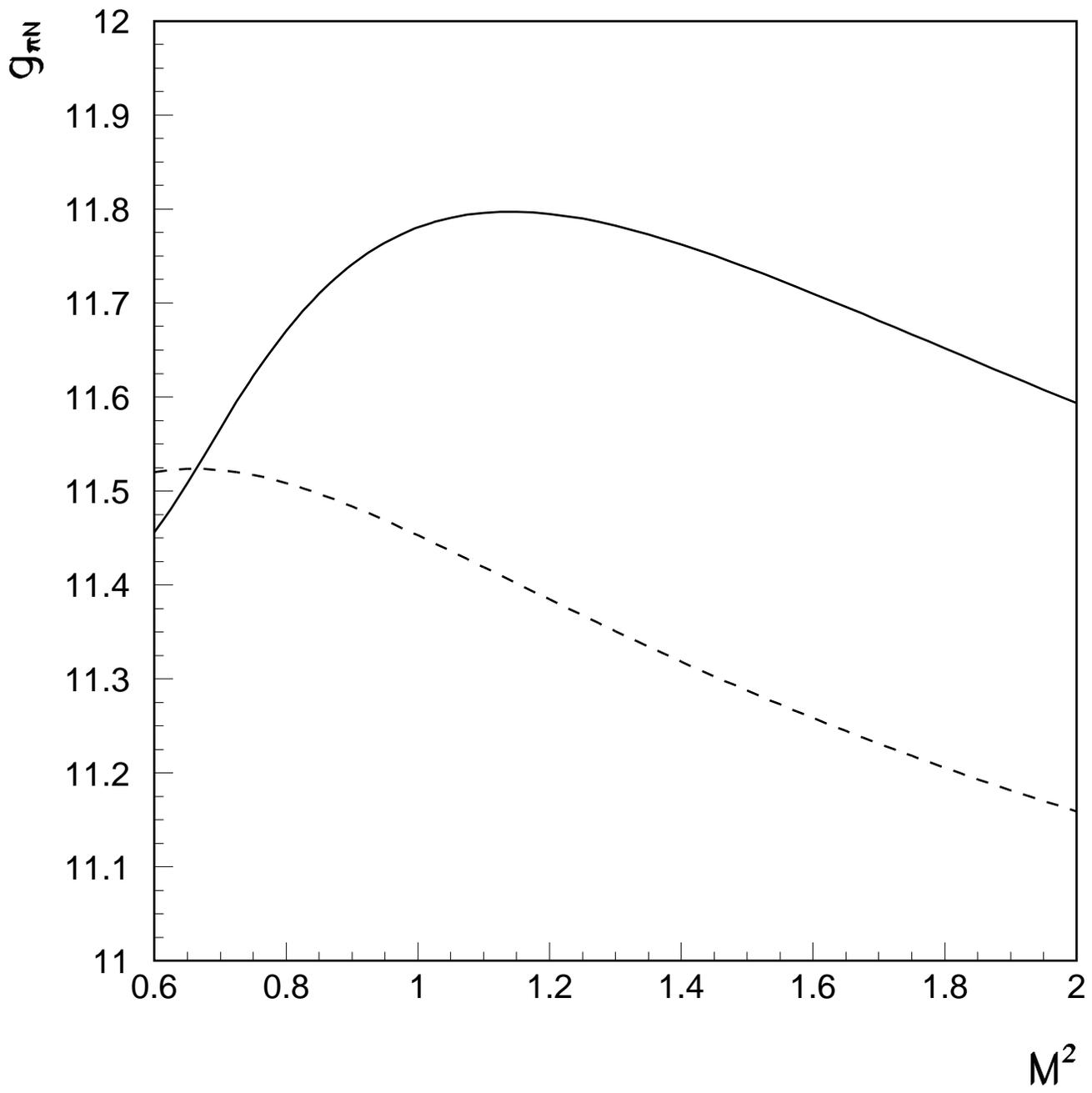
\end{document}